\documentclass[conference]{IEEEtran}
\usepackage{booktabs}
\usepackage{url}
\usepackage{cite}
\usepackage{fancyhdr}
\newcommand{\ra}[1]{\renewcommand{\arraystretch}{#1}}

\usepackage{amsmath,amssymb,amsfonts}
\usepackage{algorithmic}
\usepackage{graphicx}
\usepackage{textcomp}
\usepackage{xcolor}
\def\BibTeX{{\rm B\kern-.05em{\sc i\kern-.025em b}\kern-.08em
    T\kern-.1667em\lower.7ex\hbox{E}\kern-.125emX}}
\begin{document}

\title{Cross-Corpora Language Recognition: A Preliminary Investigation with Indian Languages}

\author{\IEEEauthorblockN{Spandan Dey$^1$, Goutam Saha$^1$, Md Sahidullah$^2$}
\IEEEauthorblockA{$^1$Dept. E\&ECE, Indian Institute of Technology, Kharagpur, West Bengal, India\\
$^2$Universit\'{e} de Lorraine, CNRS, Inria, LORIA, F-54000, Nancy, France}
\url{sd21@iitkgp.ac.in, gsaha@ece.iitkgp.ac.in, md.sahidullah@inria.fr}}

\maketitle

\begin{abstract}
In this paper, we conduct one of the very first studies for cross-corpora performance evaluation in the spoken language identification (LID) problem. Cross-corpora evaluation was not explored much in LID research, especially for the Indian languages. We have selected three Indian spoken language corpora: IIITH-ILSC, LDC South Asian, and IITKGP-MLILSC. For each of the corpus, LID systems are trained on the state-of-the-art time-delay neural network (TDNN) based architecture with MFCC features. We observe that the LID performance degrades drastically for cross-corpora evaluation. For example, the system trained on the IIITH-ILSC corpus shows an average EER of 11.80 \% and 43.34 \% when evaluated with the same corpora and LDC South Asian corpora, respectively. Our preliminary analysis shows the significant differences among these corpora in terms of mismatch in the long-term average spectrum (LTAS) and signal-to-noise ratio (SNR). Subsequently, we apply different feature level compensation methods to reduce the cross-corpora acoustic mismatch. Our results indicate that these feature normalization schemes can help to achieve promising LID performance on cross-corpora experiments.

\end{abstract}

\thispagestyle{fancy}
\fancyhf{}
\renewcommand{\headrulewidth}{0pt}
\chead{\small Accepted for publication in EUSIPCO 2021: European Signal Processing Conference, Dublin, Ireland}
\pagestyle{empty}
\lfoot{\small \copyright 2021 IEEE. Personal use of this material is permitted. Permission from IEEE must be obtained for all other uses, in any current or future media, including reprinting/republishing this material for advertising or promotional purposes, creating new collective works, for resale or redistribution to servers or lists, or reuse of any copyrighted component of this work in other works.}


\section{Introduction}
\label{sec:intro}

Voice assistants and smart devices are becoming a part of our daily life. Various speech processing applications, such as \emph{speech recognition}, \emph{speech translation}, \emph{speech synthesis}, and \emph{speaker recognition}, are very important component of these devices~\cite{ambikairajah2011language}. These speech-based applications should have a front-end \emph{language identification} (LID) module to operate on multiple spoken languages efficiently. This module can predict the spoken language from the speech input and accordingly adapt the mode of operation. 

In the last few decades, numerous significant attempts have been made to develop efficient LID systems. Different acoustic, prosodic~\cite{ng2013spoken}, ASR bottleneck features \cite{ferrer2015study,fer2017multilingually} have been utilized along with several kinds of state-of-the-art classifiers, such as, GMM, \emph{i-vector backends}~\cite{dehak2010front}, and \emph{deep neural network} based models~\cite{gonzalez2015frame,snyder2018spoken,padi2019attention}. These systems are studied mostly using a single corpus. The test data for evaluating these systems comes from the non-overlapping subsets of the same corpora used for system training. This type of evaluation does not consider the \emph{cross-corpora} variations, indicating a lack of generalization study for deploying in real-world applications. To the best of our knowledge, cross-corpora study has not been conducted explicitly for spoken language recognition. 

\begin{figure}[!t]
    \centering
    \includegraphics[trim=9cm 2cm 9cm 2cm,clip, width=.3\textwidth]{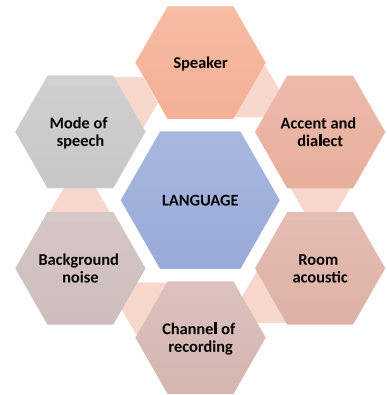}
    \vspace{-0.3cm}
    \caption{Sources of potential corpora-dependent information in LID process.}
    \label{fig:lid_variables}
    \vspace{-0.45cm}
\end{figure}

Although the cross-corpora study has not been explored explicitly in LID, researchers conducted cross-corpora study in some other speech processing applications, such as anti-spoofing~\cite{paul2017} and speech emotion recognition~\cite{schuller2010cross}. These works have found that the recognition performance degrades significantly when evaluated with audio-data from other corpora. The main reason for this is the corpora-dependent bias due to differences in data collection methods. Achieving good cross-corpora performance has remained a challenging task. In Fig.~\ref{fig:lid_variables}, we have shown several factors which can vary considerably across different corpora, collected in different settings for LID task. 

In this work, we have conducted one of the very first cross-corpora performance analyses for spoken language recognition with three standard speech corpora in Indian languages. India is a culturally and linguistically diverse country with 1.4~billion population and 22 official languages \footnote{\url{https://censusindia.gov.in/2011-common}}. For verbal interaction with the smart devices, the major portion of the Indian population is more comfortable with their respective native languages rather than English or other global languages. Due to the mutual influence and similarity among the Indian languages~\cite{sengupta2015study}, developing Indian LID systems has its unique challenges~\cite{maity2012iitkgp}. Considering these facts, researchers have given special attention to build efficient LID systems~~\cite{nandi2015implicit,dutta2018language,bhanja2019pre,mounika2016investigation,mandava2019investigation} for the Indian languages. However, these systems were developed mostly using a single corpus.

We have trained three independent LID systems with the three used corpora's training data and evaluated each system with test data from all three corpora. We have shown that the cross-corpora performances are severely inferior as compared to the same-corpora performance. Further, we have analyzed various factors due to which \emph{non-lingual} biases can be present in the individual corpus. Finally, we use various feature-level compensation methods to reduce the corpora mismatch, which substantially improves cross-corpora generalization.

\section{Corpora Description}
\label{sec:format}

We have used three standard datasets for the cross-corpora evaluation, which are widely used in the Indian LID research. These are IIITH-ILSC (IIITH)~\cite{vuddagiri2018iiith}, LDC South Asian (LDC)~\cite{LDC}, and IITKGP-MLILSC (KGP)~\cite{maity2012iitkgp}. The comparative description of the three corpora is given in Table~\ref{tab:database_compare}. For our experiments, we have chosen five languages that are common to all the corpora. These are: \emph{Bengali}, \emph{Hindi}, \emph{Punjabi}, \emph{Tamil}, and \emph{Urdu}. All the speech segments are converted into 8~kHz sampling rate, and silence regions are removed before further processing. The IIITH and KGP corpora are already divided into training and testing parts. We have manually split the data into training and testing parts with $80:20$ ratio for LDC. For all the corpora, the speakers in training and testing data are disjoint. 
\begin{table}[t]
\caption{Description of the three corpora used in this study.}
\label{tab:database_compare}
\centering
\renewcommand{\arraystretch}{1.3}
%\resizebox{.95\textwidth}{!}{%
\centering
\begin{tabular}{@{} l l l l @{} }
\toprule
\textbf{Corpora}                     & \textbf{IIITH}            & \textbf{LDC} & \textbf{KGP} \\ 
\midrule
\textbf{\#Languages}              & 23                             & 5                        & 27                     \\ 
%\textbf{Speakers/Language} & 50                             & 110                      & 10                     \\ 
\textbf{Mode of speech}    & BN and CTS              & CTS                      & BN              \\ 
\textbf{Environment}             & Studio, real-world & Real-world                & Studio        \\ 
\textbf{Total speakers}         & 1150                           & 584                & 300                    \\ 
\textbf{Duration}       & 103.5 hours                    & 118.3 hours              & 27 hours               \\ 
\textbf{Audio format}                  & 16~kHz (.wav)                   & 8~kHz (.flac)             & 8~kHz (.wav)            \\ 
\bottomrule
\end{tabular}%
\vspace{-0.3cm}
\end{table}
From the available metadata, we can summarize the major differences across the used corpora as follows:

\begin{itemize}
    \item LDC corpus contains conversational telephone speech (CTS), KGP corpus contains broadcast news (BN) data, and IIITH corpus contains both.
    
    \item In IIITH and KGP, languages are spoken in the standard form. Whereas in LDC, the speakers use local dialects and accents during conversation.
    
    \item The room environment is recording studio for KGP data. Whereas, for LDC data, no clear conclusion can be made. IIITH contains recording studio, office room, and outdoor environments.
    
    \item Variations in background noise levels are much more in IIITH data, moderate in KGP data, and less in LDC.
    
\end{itemize}

\section{Analysis of the speech corpora}
\label{sec:analysis}

To investigate the differences among the speech corpora, we have analyzed the three corpora by comparing the \emph{long-term average spectrum} (LTAS) and the \emph{signal-to-noise ratio} (SNR) histogram. We have chosen these two attributes as they represent fundamental characteristics of audio-data.

\subsection{SNR histogram}
The SNR histogram helps to analyze the level of background noise present across the speech segments in a speech corpora. We use the NIST STNR tool to calculate the overall signal-to-noise ratio corresponding to each of the speech segments\footnote{\url{https://www.nist.gov/itl/iad/mig/tools}}. 

\begin{figure}[htbp]
    \centering
    \vspace{-0.3cm}
    \includegraphics[width=.49\textwidth]{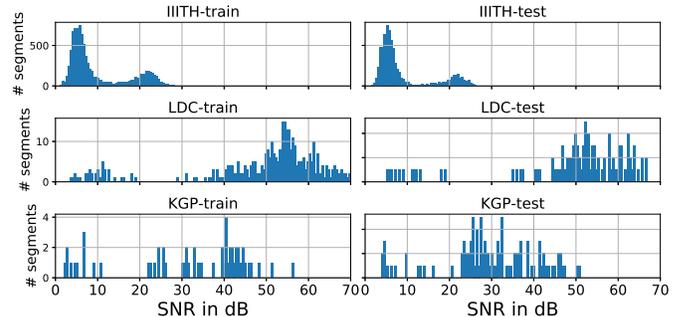}
    \vspace{-0.7cm}
    \caption{Comparison of SNR histograms for the segments in each corpora.}
    \label{fig:snr}
    \vspace{-0.1cm}
\end{figure}

In Fig.~\ref{fig:snr}, histograms of the signal-to-noise ratio values for each of the corpora are shown. Majority of the segments from IIITH corpus have SNR values less than 10 dB, with some segments having SNR values around 20 dB. In contrast, most LDC speech segments have SNR higher than 50~dB, with few low SNR segments. These facts can be justified by the bi-modal nature of the histograms for IIITH and LDC corpora. The histogram for KGP corpus is spread across a wide range of SNR values, indicating more variations in the level of background noise.

\begin{figure}[!t]
    \centering
    \includegraphics[trim=0.8cm 0.2cm 0.8cm 0.2cm, clip,width=.5\textwidth]{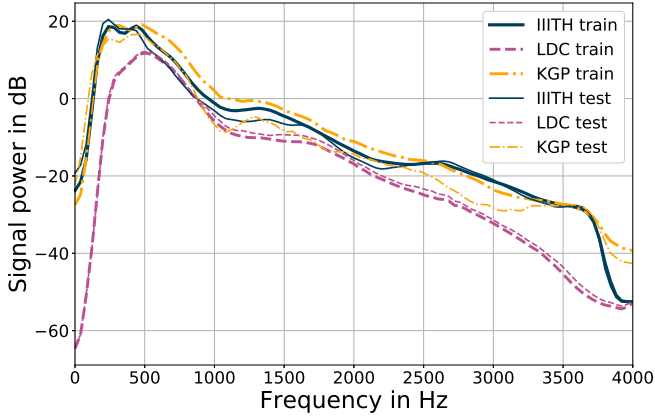}
    \vspace{-0.5cm}
    \caption{Comparison of overall LTAS of the corpora.}
    \vspace{-0.5cm}
    \label{fig:ltas_commp}
\end{figure}

\subsection{LTAS analysis}
We have compared the corpora in terms of overall spectral information using LTAS analysis. After silence removal, for each segment, the short-time Fourier transform (STFT), $S(f,t) \in \mathbb{R}^{(F \times T)}$ is computed. Here, $F$ denotes the number of frequency bins, and $T$ denotes the time frames. Then, power spectrum $P(f,t)$ is calculated from the STFT output. The LTAS spectrum for a segment ($L_s(f,t)$) is calculated by taking the log of the power spectrum:
\begin{equation}
    L_s(f,t)=\frac{1}{T}\sum_{t=1}^T \log |S(f,t)|^2
    \label{eq:ltas}
\end{equation}
Then, overall LTAS for a corpus is computed by taking the average across all its segments. In Fig.~\ref{fig:ltas_commp}, the LTAS plots are shown for the training and testing parts of the three corpora. We conclude that the spectral information among the corpora differs significantly at specific frequency ranges, mostly below 100 Hz and above 1~kHz. These ranges of frequency components mainly contribute to the non-speech factors \cite{kim2016power}, such as channels, background noise, etc. 
 
\section{Methods for mismatch reduction}
\label{sec:compensation}
The analysis conducted in Sec.~\ref{sec:analysis} shows that the mismatch among the corpora is mainly due to varying channel effects and background noise. This section discusses several feature compensation schemes to reduce the impact of those variations and apply the same in this work. Among several compensation techniques, we have applied some of the most commonly used ones in the literature, such as CMVN, feature warping, RASTA filtering, along with the relatively newer PCEN approach. Our objective is to prove the effectiveness of these popular techniques for achieving better cross-corpora generalization.

\subsection{Cepstral mean and variance normalization (CMVN)}
CMVN assumes that the channel to be static throughout the entire utterance. Thereby, it provides channel-compensated features by making mean zero with mean subtraction followed by variance unity with scaling by the inverted standard deviation. If only mean is subtracted, then this method is called cepstral mean subtraction (CMS). CMVN is expressed as,
\begin{equation}
    \mathbf{\hat{X}}=\frac{\mathbf{X}-\mathbf{\mu}}{\mathbf{\sigma}}
    \label{eq:cmvn}
\end{equation}
where, $\mathbf{X} \in \mathbb{R}^{N_d \times T}$ is the feature matrix for an utterance with $N_d$ dimensions and $T$ frames. $\mathbf{\mu},\mathbf{\sigma} \in \mathbb{R}^{N_d \times 1}$ are the mean and standard deviation respectively across the dimensions. We have also applied windowed CMVN (W-CMVN), where CMVN is applied over windows of three seconds in the utterance. In W-CMVN, the channel is assumed to remain static for a smaller duration, which is a realistic assumption.

\subsection{Feature warping (FW)}
\emph{Feature warping} (FW) warps the short-term distribution of the cepstral features to a standardized distribution~\cite{pelecanos2001feature}. FW increases feature robustness against slowly varying additive noise and various channel and transducer mismatches. 

\subsection{Relative spectral (RASTA) processing}
RASTA~\cite{hermansky1994rasta} acts as a bandpass filter and removes the static and slowly varying channel effects present in the speech signal. In the high-frequency regions, this bandpass filter also reduces the effect of convolutional noise. Hence, RASTA improves the robustness against environmental variations present in the data. The filter transfer function \cite{hermansky1994rasta} is expressed as,
\begin{equation}
    H(z) = 0.1*z^4 * \frac{2+z^{-1}-z^{-3}-2z^{-4}}{1-0.98z^{-1}}
\end{equation}

\subsection{Per-channel energy normalization (PCEN)}
We have applied PCEN~\cite{lostanlen2018per} on the mel-spectrograms ($E(f,t)$). During mel-spectral processing, PCEN replaces the logarithmic compression with the following stages~\cite{wang2017trainable}: temporal integration~(\ref{eq:integration}), adaptive gain control (AGC)~(\ref{eq:agc}), and finally dynamic range compression (DRC)~(\ref{eq:drc}),
\begin{equation}
    M(f,t) = (E * \phi_T)(t,f)=sE(t,f)+(1-s)M(t-\tau,f)
    \label{eq:integration}
\end{equation}
\begin{equation}
    G(t,f) = \frac{E(t,f)}{(M(t,f)+\epsilon)^\alpha}
    \label{eq:agc}
\end{equation}
\begin{equation}
    PCEN(t,f) = (G(t,f)+\Delta)^r-\Delta^r
    \label{eq:drc}
\end{equation}
Where, $\phi_T(t)$ is a first-order IIR filter with $0<s<1$ and $\tau$ as hop size, $0<\alpha<1$, $\Delta>1$, and $r>0$. PCEN increases the robustness against stationary background noise as well against foreground loudness variations \cite{wang2017trainable}. 

\begin{table*}[t]
\scriptsize
\centering
\caption{Baseline results for cross-corpora language recognition.}
\label{tab:baseline}
\ra{1.1}
\begin{tabular}{@{}rrrrcrrrcrrr@{}}
%\toprule
& \multicolumn{3}{c}{Training: IIITH} & \phantom{abc}& \multicolumn{3}{c}{Training: LDC} &
\phantom{abc} & \multicolumn{3}{c}{Training: KGP}\\
\cmidrule{2-4} \cmidrule{6-8} \cmidrule{10-12}
& IIITH-test & LDC-test & KGP-test && IIITH-test & LDC-test & KGP-test && IIITH-test & LDC-test & KGP-test\\ \midrule
Duration & EER / $C_{avg}$ & EER / $C_{avg}$ & EER / $C_{avg}$ & & EER / $C_{avg}$ & EER / $C_{avg}$ & EER / $C_{avg}$ & & EER / $C_{avg}$ & EER / $C_{avg}$ & EER / $C_{avg}$\\
\midrule
3~sec & 12.69 / 14.51 & 43.39 / 46.23 & 46.26 / 43.18 && 44.65 / 46.66  & 28.59 / 29.31 & 46.58 / 48.00 && 59.80 / 50.00 & 52.13 / 46.05 & 12.33 / 12.39\\
6~sec & 11.65 / 13.30 & 43.39 / 46.00 & 45.17 / 42.88 && 44.75 / 46.36 & 26.73 / 28.30 & 45.92 / 44.40 && 58.56 / 50.00 & 53.01 / 45.69 & 11.22 / 10.28\\
9~sec &11.07 / 13.19 &43.24 / 45.89 & 45.00 / 42.57 && 45.09 / 45.84 & 26.27 / 27.98 & 45.88 / 46.50 && 59.53 / 50.00 & 58.19 / 45.68 & 10.00 / ~9.63 \\
\bottomrule
\end{tabular}
\end{table*}

\begin{table*}[t]
\scriptsize
\centering
\caption{Results in terms of EER in \% / $C_{avg} \times 100$ for different feature compensation methods. M0 indicates CMS, M1 indicates CMVN, M2 indicates W-CMVN, M3 indicates FW, M4 indicates RASTA and M5, indicates PCEN.}
\label{tab:cross-corpora}
\vspace{-0.3cm}
\ra{1.1}
\begin{tabular}{@{}rrrrcrrrcrrr@{}}
%\toprule
& \multicolumn{3}{c}{Training: IIITH} & \phantom{abc}& \multicolumn{3}{c}{Training: LDC} &
\phantom{abc} & \multicolumn{3}{c}{Training: KGP}\\
\cmidrule{2-4} \cmidrule{6-8} \cmidrule{10-12}
& IIITH-test & LDC-test & KGP-test && IIITH-test & LDC-test & KGP-test && IIITH-test & LDC-test & KGP-test\\ \midrule
%Duration & EER/ $C_{avg}$ & EER/ $C_{avg}$ & EER/ $C_{avg}$ & & EER/ $C_{avg}$ & EER/ $C_{avg}$ & EER/$C_{avg}$ & & EER/ $C_{avg}$ & EER/ $C_{avg}$ & EER/$C_{avg}$\\
%\midrule
\multicolumn{10}{l}{\underline{\textbf{Test duration: 3~sec}}}\\
M0 & 12.40 / 13.82 & \textbf{43.10} / 46.23 & \textbf{30.30 / 29.91} && 50.08 / 47.34  & \textbf{22.95 / 26.40} & 45.60 / 45.92 && 37.55 / \textbf{34.91} & 51.80 / 46.72 & 11.18 / 10.82\\
M1 & 11.49 / 12.91 & 43.11 / 46.83 & 35.99 / 35.82 && 43.75 / 42.37 & 23.41 / 27.81 & 41.45 / \textbf{39.81} && \textbf{35.78} / 35.79 & 51.32 / 48.72 & 14.64 / 15.75 \\
M2 & 12.40 / 14.42 & 45.12 / 46.98 & 37.06 / 35.99 && 39.75 / 39.81 & 23.68 / 27.75 & 41.45 / 42.41 && 42.49 / 43.69 & 51.95 / 49.36 & 14.06 / 15.10 \\
M3 & 11.05 / 12.21 & 43.81 / 46.37 & 38.83 / 38.08 && 38.40 / \textbf{38.39} & 23.47 / 27.96 & 43.54 / 43.22 && 45.47 / 43.06 & 51.44 / 50.00 & 14.04 / 13.76 \\
M4 & \textbf{9.77 / 11.42} & 45.23 / \textbf{45.17} & 33.09 / 33.06 && 47.35 / 45.36 & 23.40 / 28.63 & 41.23 / 42.31 && 36.84 / 35.77 & \textbf{48.48 / 44.55} & \textbf{10.47 / 10.32} \\
M5 & 14.13 / 15.77 & 48.83 / 45.39 & 41.14 / 40.35 && \textbf{34.59} / 38.46 & 26.04 / 29.68 & \textbf{38.86} / 40.10 && 38.41 / 39.78 & 51.71 / 47.70 & 15.00 / 14.24 \\
\multicolumn{10}{l}{\underline{\textbf{Test duration: 6~sec}}}\\
M0 & 11.18 / 12.52 & 42.12 / 45.64 & \textbf{29.95 / 28.35} && 49.18 / 46.25 & \textbf{20.91 / 24.37} & 43.88 / 44.10  && 36.17 / 34.50 & 50.65 / 46.44 & \textbf{9.18} / ~8.50\\
M1 & 10.41 / 11.56 & \textbf{41.88} / 46.11 & 33.98 / 34.81 && 42.92 / 41.63 & 21.28 / 25.11 & 40.88 / 38.24 && \textbf{34.58} / 34.59 & 51.14 / 47.94  & 11.67 / 11.22 \\
M2 & 11.25 / 12.67 & 44.68 / 46.30 & 33.18 / 33.14 && 39.53 / 39.42  & 21.42 / 25.07 & 42.86 / 42.02 && 40.98 / 42.39 & 51.76 / 48.82 & 12.25 / 11.86 \\
M3 & 9.74 / \textbf{10.87} & 42.73 / 45.58 & 37.54 / 36.99 && 37.65 / 37.68 & 21.13 / 25.42 & 43.88 / 41.83 && 44.32 / 42.41 & 51.22 / 49.03 & 12.25 / 11.71 \\
M4 & \textbf{9.05} / 11.42 & 44.76 / \textbf{43.80} & 31.80 / 33.11 && 45.46 / 44.44 & 21.29 / 25.47 & 37.76 / 38.75 && 35.58 / \textbf{34.46} & \textbf{48.42 / 44.12} & \textbf{9.18 / ~8.40}\\
M5 & 12.38 / 13.38 & 43.95 / 44.45 & 37.24 / 39.05 && \textbf{33.41 / 36.59} & 23.61 / 25.91 & \textbf{36.41 / 36.67} && 37.84 / 38.99 & 52.01 / 47.45 & 13.59 / 12.41 \\
\multicolumn{10}{l}{\underline{\textbf{Test duration: 9~sec}}}\\
M0 & 10.68 / 12.25 & 42.08 / 45.28 & \textbf{32.00 / 27.63} && 50.77 / 46.88 & 20.02 / \textbf{23.36} & 44.00 / 43.25 && 37.07 / 34.17 & 49.91 / 46.23 & 8.13 / ~8.63 \\
M1 & 9.73 / 11.28 & \textbf{41.78} / 43.97 & 36.75 / 33.63 && 44.49 / 41.08 & \textbf{19.68} / 24.32 & 41.00 / \textbf{37.75} && \textbf{35.28 / 33.74} & 51.09 / 46.70 & 12.00 / 11.38 \\
M2 & 10.92 / 12.75 & 44.34 / 46.07  & 37.00 / 33.88 && 40.94 / 38.96 & 20.10 / 24.12 & 44.00 / 42.00 && 41.83 / 42.17 & 51.83 / 47.97 & 12.87 / 12.00 \\
M3 & 9.41 / 10.78 & 42.55 / 45.26 & 39.88 / 36.75 && 39.11 / 37.58 & 19.73 / 24.09 & 44.00 / 41.75 && 44.68 / 42.07 & 51.55 / 48.27 & 12.00 / 11.38 \\
M4 & \textbf{8.70 / 10.21} & 44.01 / \textbf{42.93} & 34.00 / 32.50 && 46.44 / 44.14 & 20.18 / 24.24 & \textbf{39.00} / 38.88 && 36.13 / 34.06 & \textbf{48.29 / 44.11} & \textbf{8.00 / ~7.88} \\
M5 & 12.23 / 13.22 & 43.39 / 44.21 & 41.00 / 39.75 && \textbf{33.56 / 36.61} & 21.97 / 24.45 & \textbf{39.00 / 37.75} && 37.82 / 39.05 & 51.66 / 47.24 & 11.00 / 11.00 \\
\midrule
\end{tabular}
\end{table*}

\section{Experimental setup}
\subsection{Language recognition system}
We have computed 20-dimensional \emph{mel-frequency cepstral coefficients} (MFCCs) features from the silence removed speech using 20~ms frame-size and 10~ms overlap. For classification, we have implemented \emph{time-delay neural network} (TDNN) based architecture as described in~\cite{snyder2018spoken} using the PyTorch library \cite{paszke2019pytorch}. This architecture has five convolutional layers, followed by statistical pooling, and three fully connected layers. We have used categorical cross-entropy as the loss function and AdamW~\cite{loshchilov2018decoupled} as the optimizer with a learning rate of $0.001$. The systems are trained for 30 epochs with validation loss based early stopping criteria and patience of 3 epochs. We follow an \emph{end-to-end} approach for classification. It gives better validation performance as compared to the backend-based approaches~\cite{snyder2018x}. Using the training data of each of the three corpora, we have trained three independent LID models. Then, each model is evaluated with the test data from all three corpora. If the corpora for training and test are the same, we call this as \emph{within-corpora} evaluation condition. In contrast, \emph{cross-corpora evaluation} considers the test data from different corpora than training.  

\subsection{Performance evaluation metrics}
We have evaluated the LID performance in terms of two standard metrics: \emph{equal error rate} (EER) \cite{brummer2006application} and \emph{cost average} ($C_{\mathit{avg}}$). Using \emph{detection error trade-off} (DET) plot, the false acceptance and false rejection rates are varied by changing the threshold, and EER is found at the value for which they are equal. In NIST language recognition evaluation (LRE)~\cite{sadjadi20182017} and OLR challenge \cite{li2020ap20}, $C_{\mathit{avg}}$ was used as the primary evaluation metric. It is defined as follows \cite{sadjadi20182017}:
\begin{equation} C_{avg}=\frac{1}{N}\sum\limits_{L_{t}}\begin{Bmatrix} P_{Target}\cdot P_{Miss}(L_{t})\\ +\sum\limits_{L_{n}}P_{Non-Target}\cdot P_{FA}(L_{t}, L_{n}) \end{Bmatrix} \end{equation}
where, $L_t$ and $L_n$ are the target and non-target languages. $P_{Miss}$, $P_{FA}$ are the probability of missing and false alarm. $P_{Target}=0.5$ is the prior-probability of the target languages. $P_{Non-Target}=(1-P_{Target})/(N-1)$, where, $N$ is the total number of languages. The lower value of EER and $C_{\mathit{avg}}$ indicates better classification performance.

\section{Results}

\subsection{Cross-corpora evaluation: baseline experiment}
The evaluation results of our baseline experiment are shown in Table~\ref{tab:baseline}. The results show that the average EER across within-corpora utterances is $11.80 \%, 27.20 \%$, and $11.18 \%$ for the systems trained with IIITH, LDC, and KGP corpus respectively. The relatively higher within-corpora EER for LDC corpus indicates that this corpus contains significant non-lingual variations and is difficult to learn. Cross-corpora evaluations are severely inferior as compared to within-corpora evaluations. The reason for this is the mismatches among the corpora, aroused due to variations in channel and background noise levels, as discussed in Section~\ref{sec:analysis}. This performance gap shows that LID systems trained on a single corpus can have poor generalization if deployed in real-world scenarios where the environmental variations are even more diverse.

\subsection{Effect of feature post-processing}
\begin{figure}[!t]
    \centering
    \vspace{-.3cm}
    \hspace{-.5cm}
\includegraphics[trim=9.5cm 0cm 9.5cm .9cm,clip,width=.38\textwidth]{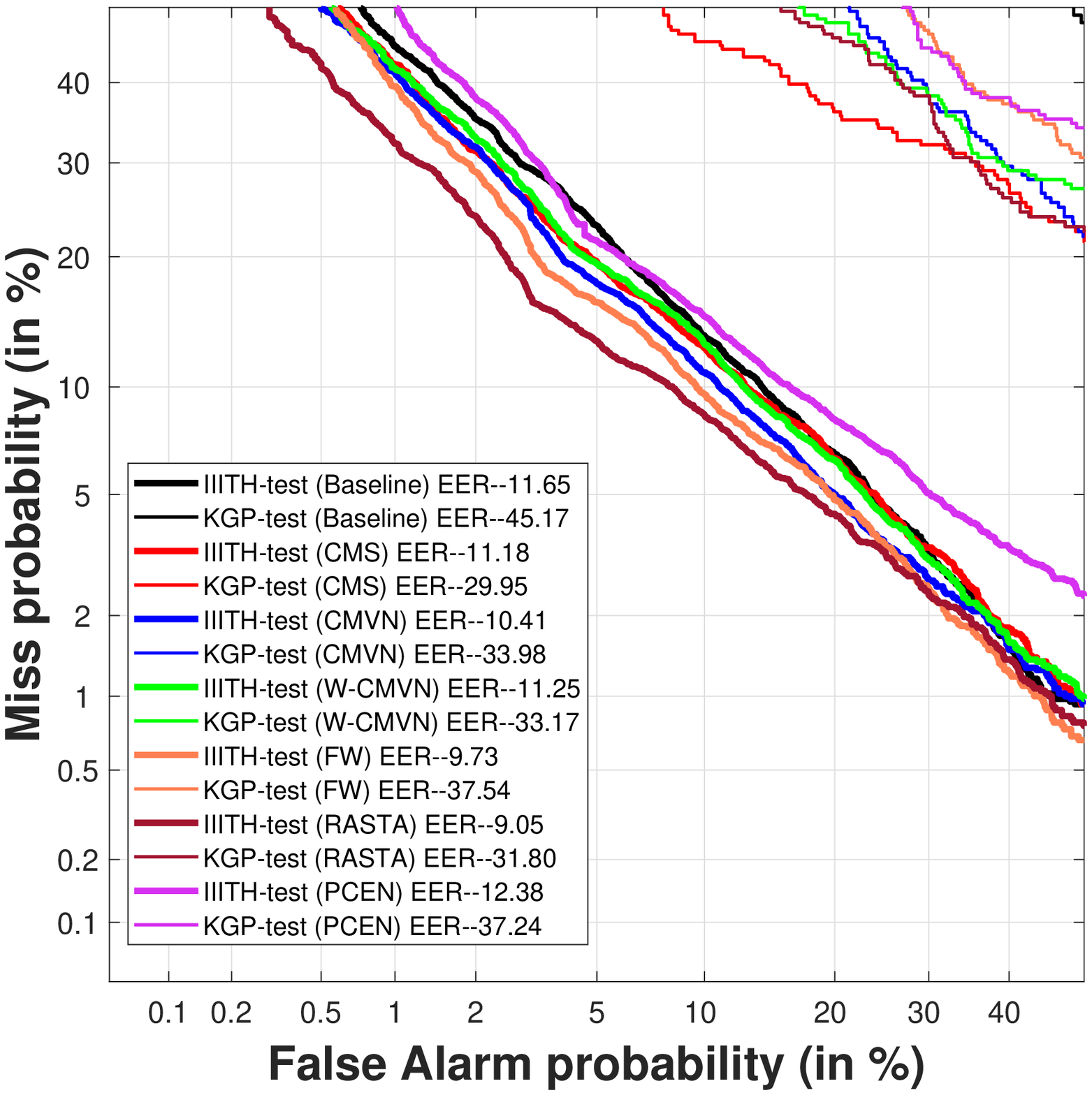}
    \vspace{-0.3cm}
    \caption{DETs of cross-corpora performances of the compensation techniques when the IIITH corpus is used in training.}
    \vspace{-0.55cm}
    \label{fig:DET}
\end{figure}

In Table~\ref{tab:cross-corpora}, we have shown the cross-corpora evaluation for the modified features. The results show that the cross-corpora performances are substantially improved as compared to the baseline performances due to the reduction in mismatch. The within-corpora performances are also improved considerably. For IIITH and KGP trained systems, RASTA processing, and for LDC trained system, PCEN achieve the best cross-corpora improvement. Generally, language recognition performance improves if the test utterance duration is increased \cite{li2013spoken,fernando2017bidirectional}. However, for our experiments, this improvement is not very prominent. The TDNN systems are trained with chunks of 3s, and the models are prone to be biased for this 3s duration \cite{raj2019probing}. In Fig.~\ref{fig:DET}, the performance comparison of the different compensation techniques for the IIITH-trained system and 6s test utterances are shown, using DET plot. The cross-corpora performance gap reduces more effectively between the IIITH and KGP corpus as compared to the LDC corpus. This is because the mismatch is higher between LDC and the other two corpora, which is also justified from Fig.~\ref{fig:ltas_commp}. Cross-corpora performance is considerably poor when the KGP-trained systems are evaluated with LDC test set. In these cases, even after the compensation, cross-corpora evaluations show  EER more than $50\%$, which is an interesting fact for further investigation. This indicates that the classifier captures the non-lingual similarities, which affect the similarity score between the target language and test audio \cite{sturm2014simple}.
\section{Conclusions}
We have studied the cross-corpora performance for spoken language recognition with three corpora. We train three independent LID systems with audio-data from each corpus using a TDNN-based language recognition system. During the evaluation, test data is used from all three corpora. In the baseline experiment, we have shown that the cross-corpora performance is almost around the chance level. We have analyzed that the environmental mismatch is one of the major causes for this. Based on the analysis, we have shown that with the feature level compensation techniques, the corpora mismatch reduces, which leads to a significant improvement in the cross-corpora performance. Among the techniques, CMS and RASTA is found to be more effective for improving generalization. The used compensation techniques can be further improved by tuning several important parameters in a learnable approach. Apart from processing at the feature level, we will also explore the effectiveness of signal level processing. We would also like to solve this cross-corpora problem from domain adaptation perspective in future study. 

%\nocite{*}
%\citestyle{authoryear}
\bibliographystyle{IEEEtran}
\bibliography{main}

\end{document}